\documentclass[letterpaper, 10 pt, conference]{ieeeconf}  % Comment this line out if you need a4paper
\title{\LARGE \textbf{
Tracking Control by the Newton-Raphson Flow: Applications to Autonomous Vehicles
}}
\author{S. Shivam, I. Buckley, Y. Wardi, C. Seatzu, and M. Egerstedt% <-this % stops a space
\thanks{
\{sshivam6, ihbuckl, yorai, magnus\}@gatech.edu; Department of Electrical and Computer Engineering, Georgia Institute of Technology, Atlanta, GA 30332, USA.}
\thanks{seatzu@diee.unica.it; Department of Electrical and Electronic Engineering, University of Cagliari,  Italy.} 
\thanks{This work was partially supported through a grant by Ford Motor Company.}}

\usepackage{graphicx}
\usepackage{amssymb}
\usepackage{amsmath}
\usepackage{amsfonts}
\usepackage{float}
\usepackage{multirow}
\newtheorem{assumption}{Assumption}

\usepackage{capt-of}
\usepackage{algorithm}
\usepackage[noend]{algpseudocode}
\usepackage{url}

\begin{document}
\maketitle

\begin{abstract}
This paper concerns applications of a recently-developed output-tracking technique to trajectory  control
of autonomous vehicles. The technique is based on three principles: Newton-Raphson flow for solving algebraic equations, output prediction, and controller speedup.
Early applications  of the technique, made to simple systems of an academic nature,  were implemented by simple algorithms requiring modest computational efforts. In contrast, this paper tests it  on
commonly-used dynamic models to see if it can handle more complex control scenarios.  Results are derived from  simulations as well as   a laboratory setting, and they indicate effective tracking convergence
despite the simplicity of the control algorithm.
\end{abstract}

\section{Introduction}

This paper considers a tracking problem where the output process of a continuous-time dynamical system has
to match, within a given tolerance, a specified target curve. We address this problem by exploring a control technique that is based on the Newton-Raphson flow for dynamically tracking the solutions of time-dependent algebraic equations.
The rationale behind the use of the Newton-Raphson flow is that it can have
stabilizing effects on the closed-loop system, and endow the controller with effective tracking with  modest computational efforts.
The control technique has been proposed in \cite{Wardi17}
and tested on various academic examples in \cite{Wardi17,Wardi18}. The objective of this paper is to test it on more challenging control problems arising
in applications to  autonomous vehicles.

The control technique that will be presented  may not be  as general as  existing
  nonlinear regulation techniques    such as the Byrnes-
Isidori regulator \cite{Isidori90} and
 Khalil's high-gain observers for output regulation \cite{Khalil98},  nor can we claim that it is more powerful.
    However, the effectiveness of these techniques is due to
 significant computational sophistication, like nonlinear
inversions and the appropriate nonlinear normal form. On the other hand, the controller described in this paper is designed  for simplicity and its implementation can be made  by a fast algorithm.  Regarding tracking applications to traffic control of autonomous vehicles,
we do not claim that our technique outperforms extant methods based on Model Predictive Control (MPC) - the most common current approach to such applications.
However, we perform the simulation testing on two specific problems that have been addressed by MPC techniques \cite{Zhou17,Borrelli07}, and  the tracking-error
results that
we obtain are no worse. Also,  our technique may be less computing-intensive than MPC (although no explicit comparison is made
in the paper)
because it requires no solutions of optimal control problems. In this, we do not claim that our technique is better than MPC; in fact,
we think that it is not as general. We only argue that it deserves a further study for potential use in future applications.

\begin{figure}	[b]
\begin{center}
\vspace{8pt}
	\includegraphics[width=0.85\linewidth]{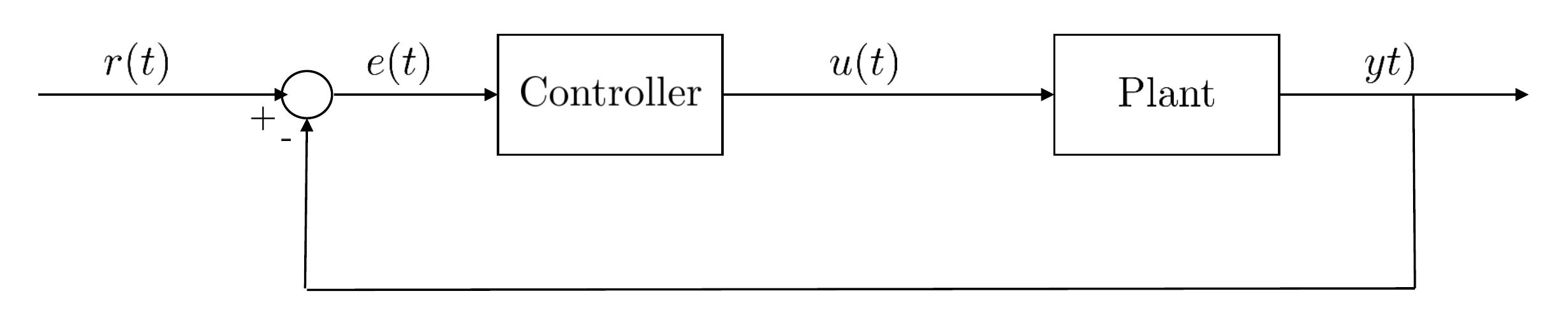}
	\vspace{-8pt}	
	\captionof{figure}{{ Basic control system. }}
\end{center}
\end{figure}

%\begin{figure}
%	\centering
%	\includegraphics[width=3.0 in]{Figure_1.pdf}
%	\caption{{ Basic control system}}
%\end{figure}

To illustrate the Newton-Raphson flow, consider the continuous-time system in Fig.~1, where ${\{r(t):t\geq 0\}}$ is a given target curve in
$R^m$, $\{u(t):t\geq 0\}$ is the control input to the plant, $\{y(t):t\geq 0\}$ is the system's output, $e(t):=r(t)-y(t)$ and $\{e(t):t\geq 0\}$ is the error signal.\footnote{In the future discussion we will refer to a process
$\{x(t):t\geq 0\}$ by $\{x(t)\}$ for the sake of notation's simplicity, and to distinguish it from its value at time $t$ which is denoted by $x(t)$.} We assume that the trajectories $\{r(t)\}$, $\{u(t)\}$  and $\{y(t)\}$, hence $\{e(t)\}$ are in the same Euclidean space, $R^m$.
To highlight the salient features of the Newton-Raphson  flow,  suppose for a moment that the  plant is a
memoryless nonlinearity relating $u(t)$ to $y(t)$ by the equation
\begin{equation}
y(t)=g(u(t))
\end{equation}
for a continuously-differentiable function $g:R^m\rightarrow R^m$. Define the controller by the following equation,
\begin{equation}
\dot{u}(t)=\left(\frac{\partial g}{\partial u}(u(t))\right)^{-1}\big(r(t)-g(u(t))\big),
\end{equation}
assuming  that the Jacobian matrix $\frac{\partial g}{\partial u}(u(t))$ is nonsingular for every $t\geq 0$. Since $y(t)=g(u(t))$, the
controller can be viewed  as a continuous-time flow of the Newton-Raphson method, tracking the solution of the time-dependent
equation $g(u)=r(t)$; hence, we label it the ``Newton-Raphson flow''.

We suggest the  suitability of the Newton-Raphson flow to  tracking applications in view of the fact that it is essentially an integrator,
a key element  in tracking control,  albeit
with a variable gain.
To see this point, observe that ${e(t)=r(t)-y(t)}$; hence, Eq.~(2) has the form
$\dot{u}(t)=A(t)e(t)$ with $A(t):=(\frac{\partial g}{\partial u}(u(t)))^{-1}$. If $A(t)\equiv 1,$ then
the controller would be the familiar integrator; as it stands, in Eq.~(2), we can say that it is a variable gain integrator.

With the aforementioned particular gain $A(t)$, stability of the closed-loop system is hardly an issue. To see this, define the function
$V:R^m\times R^+\rightarrow R^+$ by
\begin{equation}
V(u,t)=\frac{1}{2}||r(t)-g(u)||^2.
\end{equation}
Then it is readily seen (e.g., \cite{Wardi18}) that along trajectories in $u$, for every $t\geq 0$,
\begin{equation}
\dot{V}(t)
\leq -||r(t)-g(u(t))||\big(||r(t)-g(u(t))||-||\dot{r}(t)||\big).
\end{equation}
If the curve $\{r(t)\}$ is continuously differentiable and  bounded,  define  $\eta:=\sup\{||\dot{r}(t)||:t\geq 0\}$;
 then Eq.~(3) implies that
\begin{equation}
{\rm limsup}_{t\rightarrow\infty}||r(t)-g(u(t))||\leq\eta,
\end{equation}
provided that the Jacobian matrix $\frac{\partial g}{\partial u}(u(t))$ is nonsingluar at every $t\geq 0$. This implies the boundedness of
$\{u(t)\}$ and an upper bound, $\eta$, on the error $||r(t)-y(t)||$ (see Fig. 1). Note the special case where the target curve is a constant $r\in R^m$, then
$V(u(t))$ is a Lyapunov function  and $g(u(t))\rightarrow r$ as $t\rightarrow \infty$. In the general  case of a nonconstant
curve $\{r(t)\}$, the upper bound in the Right-Hand Side (RHS) of Eq.~(5) can be reduced by speeding up the controller in the following way:
multiply the RHS of Eq.~(2) by a constant $\alpha>1$, thereby redefining the controller as
\begin{equation}
\dot{u}(t)=\alpha\left(\frac{\partial g}{\partial u}(u(t))\right)^{-1}\big(r(t)-g(u(t))\big);
\end{equation}
then (see \cite{Wardi18})
\begin{equation}
{\rm limsup}_{t\rightarrow\infty}||r(t)-g(u(t))||\leq\eta/\alpha,
\end{equation}
as long as the Jacobian $\frac{\partial g}{\partial u}(u(t))$ is nonsingular along the trajectory $\{u(t)\}$.

We stress the point that the  boundedness   and tracking properties of the controller  require the sole assumption that the Jacobian
$\frac{\partial g}{\partial u}(u(t))$ is nonsingular for every $t>0$. As a matter of fact, these properties are   inherent in the direction defined by the Newton-Raphson flow.
Now it is well known that, in its standard discrete-time setting,  the Newton-Raphson method can be unstable (\cite{Lancaster66}), but this is due primarily to its step size and not its direction. In its flow setting the
step size is infinitesimal, which seems to circumvent this stability issue.

If the plant is a dynamical system, then  extensions of the controller,
defined in Eq.~(6) for memoryless systems,
are not straightforward. For example, $y(t)$ is no longer a function only of $u(t),$ but rather of
$\{u(\tau):\tau\leq t\}$;
hence, it is unclear what
$g(u(t))$ would mean. Moreover, as we shall see, stability of the closed-loop system
cannot be taken for granted. We address these issues by defining the controller
to be based on the following three principles:
the Newton-Raphson flow, redefined in a suitable sense;   an output predictor, which defines the time-dependent nonlinear equations that  the controller attempts to solve; and a controller speedup, which aims at stabilizing the closed-loop system in addition to reducing the tracking error.
The controller was defined in
\cite{Wardi17} and preliminary analysis and simulation results have been obtained in \cite{Wardi17,Wardi18} (surveyed below).
As mentioned earlier, the main contributions of this paper are in the study of applications arising in   autonomous vehicles through simulation and experimentation in a laboratory setting.

The rest of the paper is organized as follows. Sec. 2 recounts existing results and sets the stage for the later experiments.
Sec. 3 describes simulation results, and Sec. 4
presents a laboratory experiment. Sec. 5 concludes the paper and suggests directions for future research.

\section{Problem Formulation and Previous Results}
This section presents the controller's definition as presented in \cite{Wardi17}, and recounts preliminary
results in \cite{Wardi17,Wardi18}.

Suppose that the plant subsystem in Fig.~1 is a dynamical system defined by the differential equation
\begin{equation}
\dot{x}(t)=f(x(t),u(t))
\label{eq:system}
\end{equation}
and the output equation
\begin{equation}
y(t)=h(x(t)),
\label{eq:output}
\end{equation}
where $x(t)\in R^n$ is the state variable, $u(t)\in R^m$ is the input at time $t$, and $y(t)\in R^m$ is the output at time $t$.
The system evolves in the time-interval  $t\in[0,\infty)$, and the initial condition for Eq.~(8) is a given $x_{0}:=x(0)\in R^n$. The following assumption is made on the functions
${f:R^n\times R^m\rightarrow R^n}$ and ${h:R^n\rightarrow R^m}$:
\begin{assumption}
(i). The function $f:R^n\times R^m\rightarrow R^n$ is continuously differentiable, and for every compact set $\Gamma\subset R^m$
there exists $K>0$ such that, for every $x\in R^n$ and $u\in\Gamma$,
\begin{equation}
||f(x,u)||\leq K\big(||x||+1\big).
\end{equation}
(ii). The function $h:R^n\rightarrow R^m$ is continuously differentiable.
\end{assumption}

This assumption implies that for every bounded, piecewise-continuous input $\{u(t)\}$, and for every initial condition
$x_0\in R^n$, Eq.~(8) has a unique, continuous, piecewise continuously-differential solution
$\{x(t)\}$ on $t\in[0,\infty)$.

In order to extend the controller from the above memoryless setting (Eq.~(1)) to the current
context of dynamical systems, we must have a way of  expressing $y(t)$ as a function of $u(t)$ and perhaps other terms.   To this end, and in order to  ensure a suitable setting for the Newton-Raphson flow,   we predict, at time $t$,  the future output at time $t+T$ for a given $T>0$,
and  regard the predicted output  as a function of $x(t)$ and $u(t)$. Specifically, define the predicted state trajectory in the time interval
$\tau\in[t,t+T]$, denoted  by $\xi(\tau)$, by  the following differential
equation,
\begin{equation}
\dot{\xi}(\tau)=f(\xi(\tau),u(t)),
\end{equation}
with the initial condition $\xi(t)=x(t)$; then define  the predicted output, $\tilde{y}(t+T)$, by
\begin{equation}
\tilde{y}(t+T)=h(\xi(t+T)).
\end{equation}
Observe that  $\tilde{y}(t)$ is a function of $x(t)$ and $u(t)$, and this functional dependence is denoted by
\begin{equation}
\tilde{y}(t+T)=g(x(t),u(t)).
\end{equation}
The controller we define has the following form,
\begin{equation}
\dot{u}(t)=\alpha\left(\frac{\partial g}{\partial u}(x(t),u(t))\right)^{\!\!-1}\!\!\!\big(r(t+T)-g(x(t),u(t))\big),
\label{equation:u_update}
\end{equation}
and we note that this provides an extension of Eq.~(6)  as $g(x(t),u(t))$ depends  on $u(t)$ and $x(t)$ as well.
Eqs.~(8), (9)  and (14) together define the closed-loop system.

Generally, it is desirable to have the prediction
horizon be as small as possible, since a smaller $T$ usually results in  a smaller prediction error than a larger $T$, and this can translate into a smaller tracking error. Moreover, the effects of prediction errors on the tracking errors cannot be
attenuated by speeding up the controller via a large $\alpha$ in Eq.~(14), and hence, various applications may require a small  prediction horizon $T$. However, initial analyses of simple examples, carried out in \cite{Wardi17},  revealed that, for a given $\alpha>0,$ there exists
$T_{\alpha}>0$ such that the closed-loop system is stable for $T>T_{\alpha}$ and unstable for $T< T_{\alpha}$. In other words,
for $T$ small enough, the system is unstable, which places a roadblock on the requirement of a small prediction horizon.
However, it was also shown that
$T_{\alpha}$ is monotone decreasing in $\alpha$, and
\begin{equation}
\lim_{\alpha\rightarrow \infty}T_{\alpha}=0,
\end{equation}
which suggests  the following practical implication: first, choose $T$ small enough to ensure a small prediction error, then choose $\alpha$ large enough to guarantee
 stability of the closed-loop system.

 The aforementioned results were derived for particular simple systems, but they indicate that speeding up the controller may have two beneficial consequences:   stabilizing the system and reducing the asymptotic tracking error as in Eq.~(7). This was corroborated in \cite{Wardi17,Wardi18} by  simulation experiments on various systems including an inverted pendulum,  a platoon of mobile robots, and a network of mobile agents whose motion is coordinated by the graph Laplacian.  While some of these control problems were challenging, the  dynamic models and equations of motion of each of their constituent agents are quite simple,  typically  having the form of first-order or second-order linear systems.
 In contrast, as we said, in this paper we apply  the control technique on systems with more complicated and realistic  motion dynamics.

 We point out that the key theoretical question concerns the derivation of verifiable conditions (necessary or sufficient) for the following
 property of the closed-loop system: for every $T>0,$ there exists $\bar{\alpha}>0$ such that, for every $\alpha>\bar{\alpha}$, the closed-loop system is
 stable. We label this property the {\it $\alpha$-stability of the system}. Thus far, it has been proved for the two-dimensional linear systems analyzed in \cite{Wardi17},
 but an analysis in the general case is challenging due to the lack of closed-form expressions  for $g(x,u)$. Simulation-based evidence suggests that
 generally,
 stability of the closed-loop system implies tracking in the sense of Eq.~(7), and the derivation of verifiable sufficient conditions for various classes
 of systems is currently under way.
 Results of these theoretical investigations will be published elsewhere, while here we focus only on experimental
 verification in a particular application area.

\section{Simulation Results}
% \begin{figure}[]
%     \centering
%     \includegraphics[width=0.8\linewidth]{bicycle.pdf}
%     \caption{Caption}
%     \label{fig:my_label}
% \end{figure}

This section presents results of simulation experiments for two problems taken from the literature  on path
tracking by autonomous vehicles \cite{Zhou17,Borrelli07}. A key objective of this  study is to test the tracking-control
framework on system-models  that are considerably more realistic and complex than the simple models used in the past \cite{Wardi17,Wardi18}. To this
end we choose the bicycle model for the vehicles' motion dynamics, which is often used in studies of control of autonomous vehicles (see, e.g., \cite{Kong15}
and references therein). We adopt the  dynamic model presented in  \cite{Kong15,Borrelli07},  and summarize it in the following paragraphs.

\begin{figure}[b]
    \centering
    \includegraphics[width=0.9\linewidth]{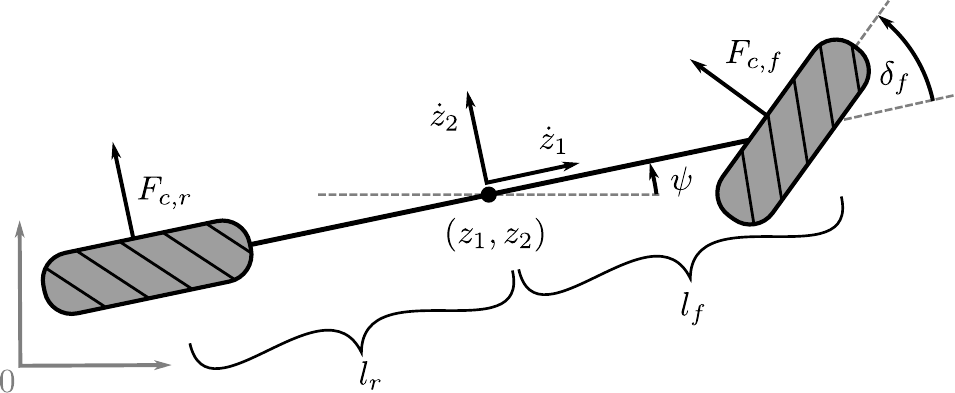}
    \caption{Dynamic bicycle model}
    \label{fig:bicycle}
\end{figure}

The state of the bicycle-model system is given by \\
${x=(z_{1},z_{2},v_{\ell},v_{n},\psi,\dot{\psi})^{\top}}$, where $z_{1}$ and $z_{2}$ are the planar coordinates of the center of gravity of the vehicle, $v_{\ell}$ is the longitudinal velocity, $v_{n}$ is the lateral velocity, $\psi$ is the vehicle's heading and $\dot{\psi}$ is its angular velocity. The control input to the bicycle model is ${u=(a_\ell , \delta_f)^\top}$, where $a_{\ell}$ is the longitudinal acceleration, and $\delta_{f}$ is the front-wheel steering angle. For clarification, an illustration of the bicycle model is included in Fig.~\ref{fig:bicycle}.

Referring to the notation used in  the previous section and especially to Eqs. (8) and (9), the state equation of the system is defined
by the following equations:
\begin{align}
\dot{z}_{1}&=v_{\ell}\cos\psi-v_{n}\sin\psi\\
\dot{z}_{2}&=v_{\ell}\sin\psi+v_{n}\cos\psi\\
\dot{v}_{\ell}&=\dot{\psi}v_{n}+a_{\ell}\\
\dot{v}_{n}&=-\dot{\psi}v_{\ell}+{2}\left(F_{c,f}\cos\delta_f+F_{c,r}\right)/{m}\\
\ddot{\psi}&={2}\left(l_f F_{c,f}\cos\delta_f-l_r F_{c,r}\right)/{I_z},
\end{align}
where $m$ is the mass of the vehicle,
$l_f$ and $l_r$ are the respective distances of the front and back axles from the vehicle's center of mass,
  $I_z$ is the yaw moment of inertia, and $F_{c,f}$ and $F_{c,r}$ are the lateral forces on the front and rear wheels.
  The forces   $F_{c,f}$ and $F_{c,r}$ are
 approximated as:
\begin{equation}
F_{c,f}=C_{\alpha,f}\left(\delta_f-\tan^{-1}\left((v_{n}+l_f\dot{\psi})/v_{\ell}\right)\right)
\end{equation}
\begin{equation}
F_{c,r}=-C_{\alpha,r}\tan^{-1}\left(({v_{n}-l_r\dot{\psi}})/{v_{\ell}}\right),
\end{equation}
where  $C_{\alpha,f}$ and $C_{\alpha,r}$ are   the cornering stiffness of the front tire and rear tire.

For the purpose of output tracking we define the  output by the planer coordinates of the center of gravity of the vehicle,
namely  $y=(z_1, z_2)^{\top}= Cx$, where
\begin{equation}
C=\left(
\begin{array}{cccccc}
1 & 0 & 0 & 0 & 0 & 0\\
0 & 1 & 0 & 0 & 0 & 0
\end{array}
\right).
\end{equation}

Instantiating the control equation   (14), the controller has to compute the terms
$g(x(t),u(t))$ and  $\frac{\partial g}{\partial u}(x(t),u(t))$ in real time, at time $t$. By Eqs. (12) and (13),
$g(x(t),u(t))=h(\xi(t+T))=C\xi(t+T)$, and $\xi(t+T)$
 can be computed
by numerical integration of the differential equation (11) in the interval $\tau\in[t,t+T]$, with the initial condition $\xi(t)=x(t)$. To this purpose we use the Forward-Euler method. As for the term $\frac{\partial g}{\partial u}(x(t),u(t))$, we observe that
\begin{equation}
\frac{\partial g}{\partial u}(x(t),u(t))=C\frac{\partial \xi(t+T)}{\partial u(t)}.
\end{equation}
Moreover, by taking derivatives with respect to $u(t)$ in Eq. (11) the following equation
for  $\frac{\partial\xi(\tau)}{\partial u(t)}$ is obtained,
\begin{equation}
\frac{d}{d\tau}\frac{\partial\xi(\tau)}{\partial{u(t)}}=\frac{\partial f}{\partial\xi}(\xi(\tau),u(t))\frac{\partial\xi(\tau)}{\partial
u(t)}+\frac{\partial f}{\partial u}(\xi(\tau),u(t)),
\end{equation}
with the  boundary condition $\frac{\partial\xi(t)}{\partial u(t)}=0$. This equation can be solved
in the interval $\tau\in[t,t+T]$ by the forward Euler method, concurrently with Eq. (8), thereby yielding
$\frac{\partial g}{\partial u}(x(t),u(t))$ via Eq. (24).

\subsection{Tracking of a closed curve}
In this subsection we test   the proposed controller on a  bicycle model having to track a given closed curve. This problem was considered in \cite{Zhou17}  and we use the same parameters. However, we use a different model, as
\cite{Zhou17} uses a kinematic model whereas we use the dynamic bicycle model described above.
The   parameters of the problem are:
$m=1,587~kg$, $I_z=2,315.3~kg~m^2$,
$l_f=1.218~m$,
$l_r=1.628~m$, and
$C_{\alpha,f}=C_{\alpha,r}=35,000~N/rad$.
 The target track is depicted in Fig.~3.
 The target moves counterclockwise along the curve at a constant reference  speed, and three simulation experiments were
 conducted  for the following corresponding
 speeds: $15~km/h$, $25~km/h$ and $35~km/h$.  The vehicle position is initialized on the track as indicated in Fig. 3 with the initial speed equal to the reference speed.
 The duration of  each  experiment is $100~s$.
  The parameters for the output-tracking controller are:  the prediction
horizon is  $T=0.5~s$, the discretization  time step
for the predictor is $\Delta t=0.0025~s$, and the  speedup coefficient in Eq. (14) is $\alpha=30$.

\begin{figure}[!ht]
\vspace{6pt}
\begin{center}
	\includegraphics[width=0.9\linewidth]{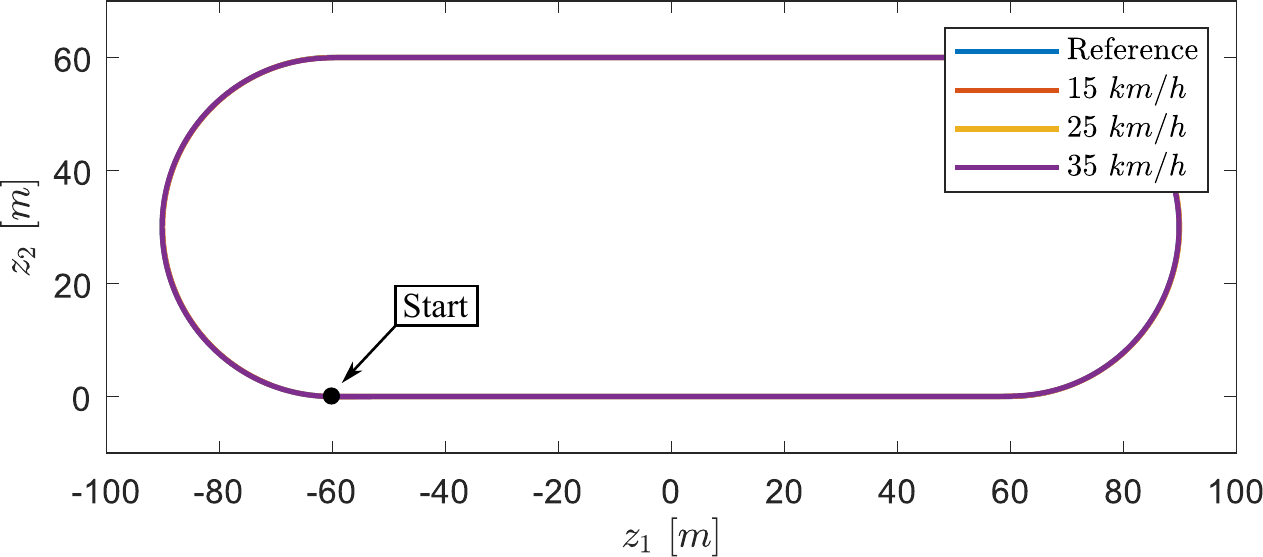}
	\vspace{-8pt}	
	\captionof{figure}{{ Closed-path tracking curve }}
\end{center}
\end{figure}
% \vspace{5pt}

\begin{figure}
\begin{center}
	\includegraphics[width=0.9\linewidth]{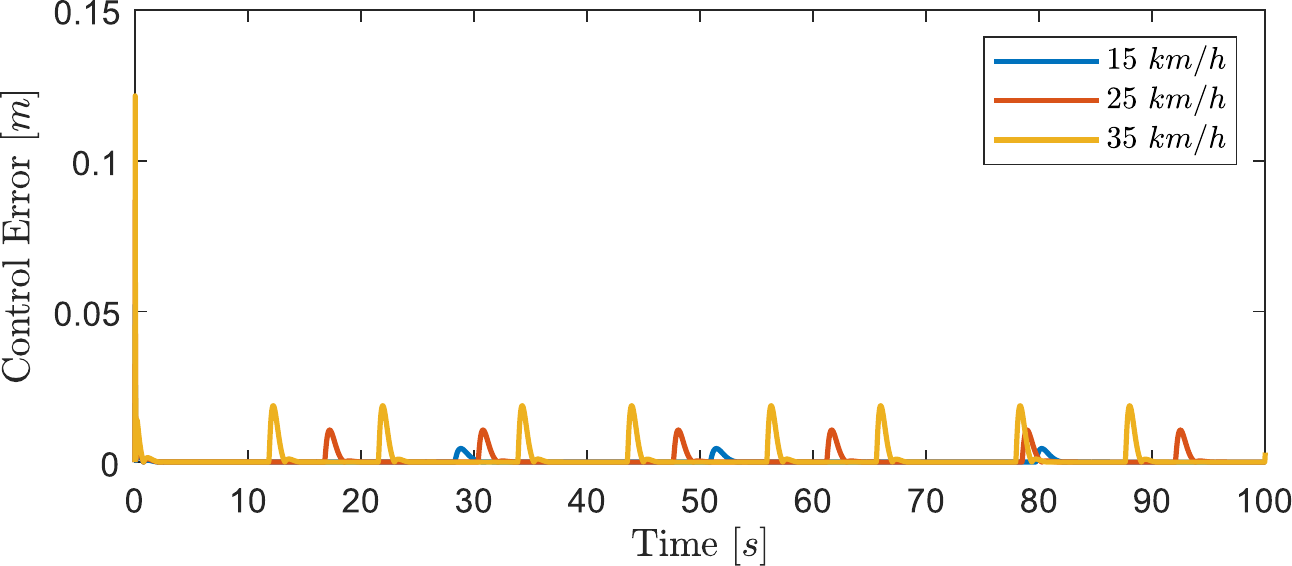}
	\vspace{-8pt}
	\captionof{figure}{{Closed path: control errors. Peak error (following an initial transient) is about $2~cm$. }}
\end{center}
\end{figure}
% \vspace{5pt}

\begin{figure}
\begin{center}
	\includegraphics[width=0.9\linewidth]{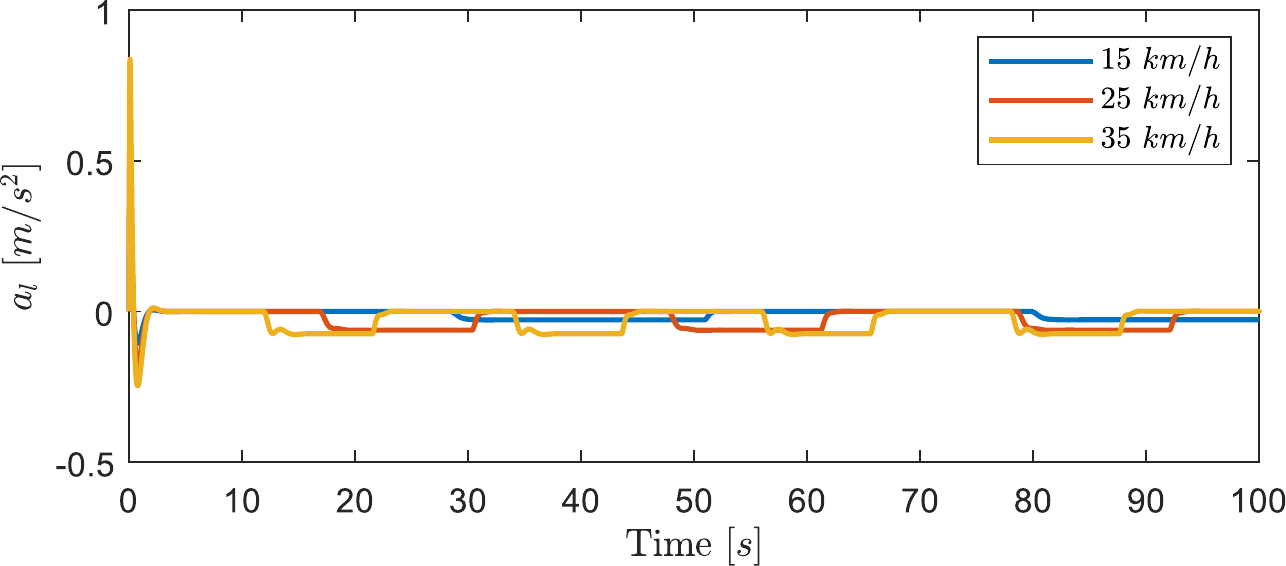}
	\vspace{-8pt}
	\captionof{figure}{{ Closed path: longitudinal acceleration. Peak acceleration (following an initial transient) is under $0.1~m/s^2$.}}
\end{center}
\end{figure}
% \vspace{5pt}

\begin{figure}
\begin{center}
	\includegraphics[width=0.9\linewidth]{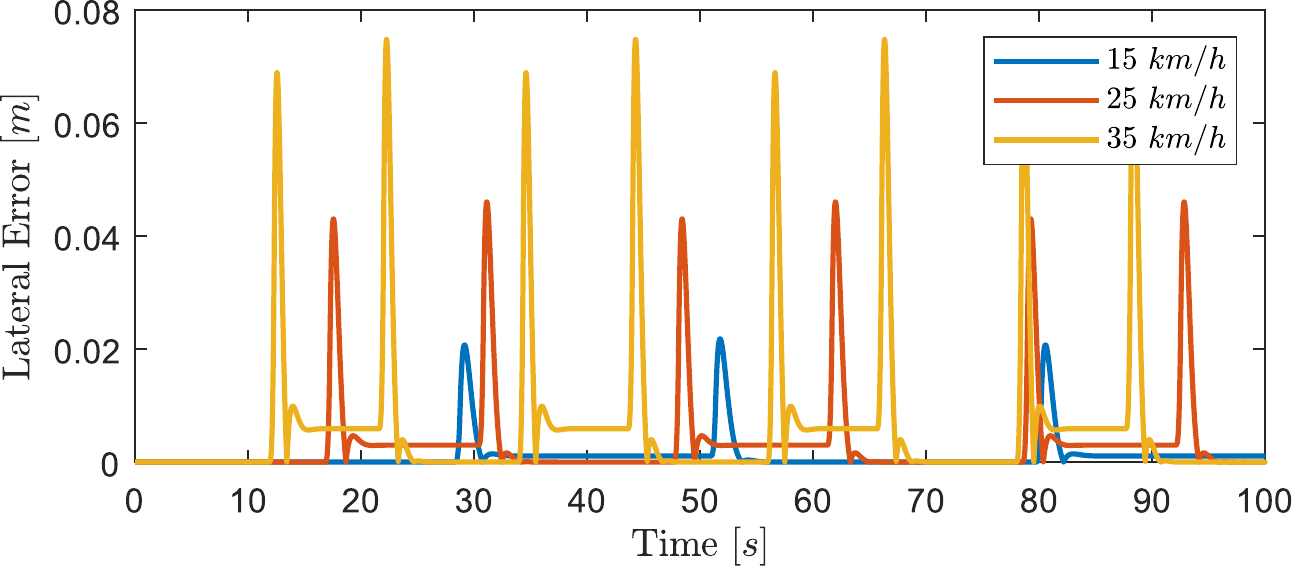}
	\vspace{-8pt}
	\captionof{figure}{{ Closed path: lateral errors. Peak error is $8~cm$.}}
\end{center}
\end{figure}
% \vspace{5pt}

\begin{figure}
\begin{center}
	\includegraphics[width=0.9\linewidth]{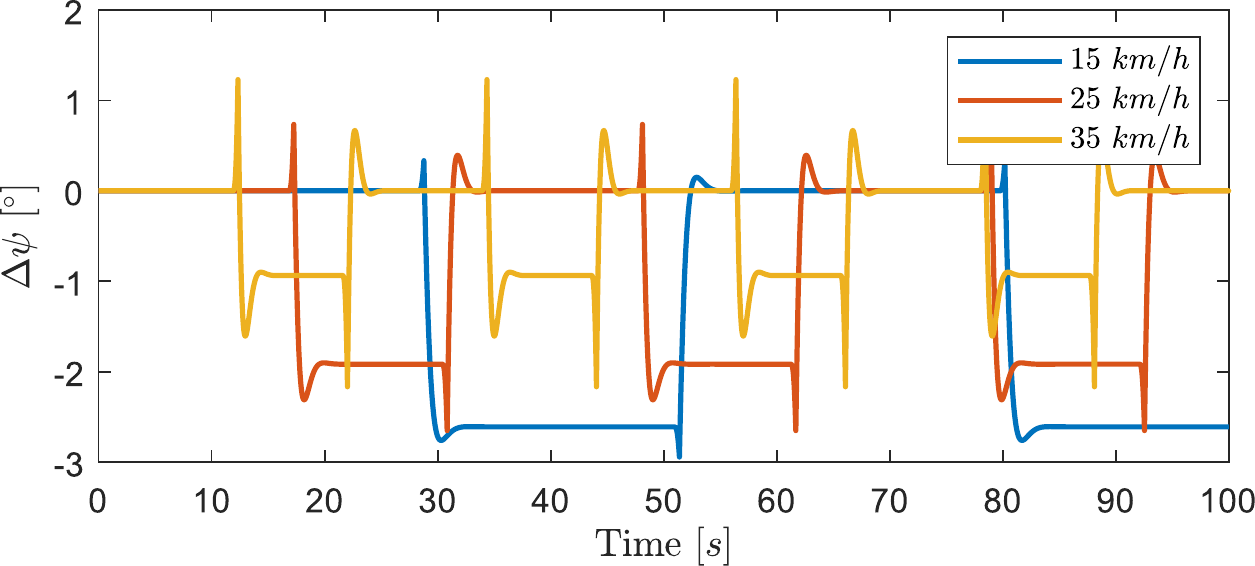}
	\vspace{-8pt}
	\captionof{figure}{{Closed path: heading errors. Peak error is under $3^{\circ}$.}}
\end{center}
\end{figure}

\vspace{5pt}
\begin{center}
\begin{small}
\begin{tabular}
{ |c| c |c | c| c|}
\hline 
\multirow{2}{4em}{Speed} & \multicolumn{2}{|c|}{Peak lateral error} & \multicolumn{2}{|c|}{Peak heading error} \\ 
\cline{2-5}
&	Shivam 19 & Zhou 17	&	Shivam 19 & Zhou 17 \\
\hline
 $15~km/h$ &	$2~cm$ 	&	$36~cm$	&	$3^{\circ}$		&	$3.3^{\circ}$ \\
 $25~km/h$ & 	$5~cm$ 	&	$23~cm$	&	$2.8^{\circ}$	&	$2.8^{\circ}$ \\
 $35~km/h$ & 	$8~cm$ 	&	$17~cm$	&	$2.2^{\circ}$	&	$2.6^{\circ}$ \\
\hline
\end{tabular}
\end{small}
\captionof{table}{Peak lateral and heading error for curve tracking\label{tab:table-one}}
\end{center}

Fig.~3  shows the reference curve as well as
  the trajectories obtained from the three experiments, and the trajectories are barely  distinguishable from the reference curve.
    Fig.~4 depicts the graphs of the  control errors, which can serve to gauge the efficacy of the control algorithm.\footnote{We define
    the control error to be be $||r(t)-\tilde{y}(t)||$, and note that it is the norm of the signal
    $e(t)$ which serves as the input to the controller subsystem in Fig. 1. A different kind of error, the tracking error, is
    defined as $||r(t)-y(t)||$.}
     
 With the exception of an initial transient,   the peak control error is about   $ 2~cm$, while counting
     the transients it is about $12.5~cm$.  The small peaks in the control error correspond with turns on the reference curve which induce
     prediction errors. Fig.~5 shows the longitudinal acceleration of the trajectories which is under $0.1~m/s^2$ except for a transient where its peak is under $0.8~m/s^2$.
Fig.~6  depicts the graphs of the lateral position errors, defined as the  distance from the vehicle's center
  of gravity to the target curve. We discern maximum errors of $2~cm$, $5~cm$, and $8~cm$ for the respective speeds of $15~km/h$, $25~km/h$, and $35~km/h$.
    The resulting heading errors are depicted in Fig.~7,
  and it is seen that the error does not exceed $3^{\circ}$.  Ref. \cite{Zhou17} also reports the lateral and heading (angular) errors obtained from its control algorithm.
  These and our simulation results are summarized in Table 1.
  
\subsection{Lane-change maneuver}
The simulation performed in this subsection is applied to the same system and problem-parameters as described in
\cite{Borrelli07}. We point out that \cite{Borrelli07} tested it by simulation and in a laboratory setting, while our results were
obtained only from simulation. Note that the Pacejka tire model was used in \cite{Borrelli07}; however, in the range of operation the lateral forces are similar to those given by Eqs. (21)-(22).

The parameters of the vehicle are  $m=2,050~kg$, $I_z=3,344~kg~m^2$, $l_f=1.105~m$, $l_r=1.738~m$, $C_{\alpha,f}=57500~N/rad$, and $C_{\alpha,r}=92500~N/rad$.
The target curve, depicted in Fig.~8,  is parameterized by the longitudinal position as given in \cite{Borrelli07}:
\begin{equation}
z_{2}=2.025\left(1+\tanh w_1\right)+2.85\left(1+\tanh w_2\right),
\end{equation}
where $w_1=(2.4/25)(z_{1}-27.19)-1.2$ and $w_2=(2.4/21.95)(z_{1}-56.46)-1.2$.  The target moves along the curve at a constant reference  speed; simulation experiments were run for the following speeds: $10~m/s$, $15~m/s$, and $19~m/s$. In each simulation the  vehicle starts at the origin at the reference speed.  The duration of  each  experiment is $25~s$. The discretization step size for the numerical
simulation is $0.01~s$.
The controller prediction horizon is  $T=0.5~s$, the discretization  time step
for the predictor is $\Delta t=0.001~s$ and the  speedup coefficient is $\alpha=30$.

\begin{figure}[!th]
\vspace{4pt}
\begin{center}
	\includegraphics[width=0.9\linewidth]{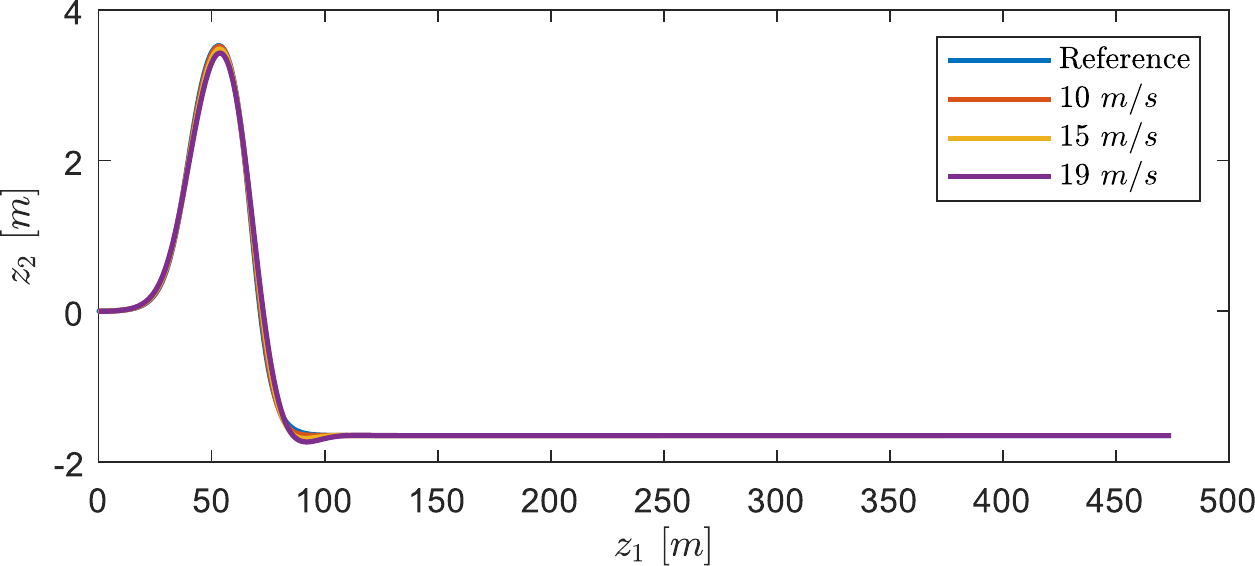}
	\vspace{-8pt}
	\captionof{figure}{{Lane-changing curve.}}
\end{center}
\end{figure}

% \vspace{5pt}
\begin{figure}
\begin{center}	
	\includegraphics[width=0.9\linewidth]{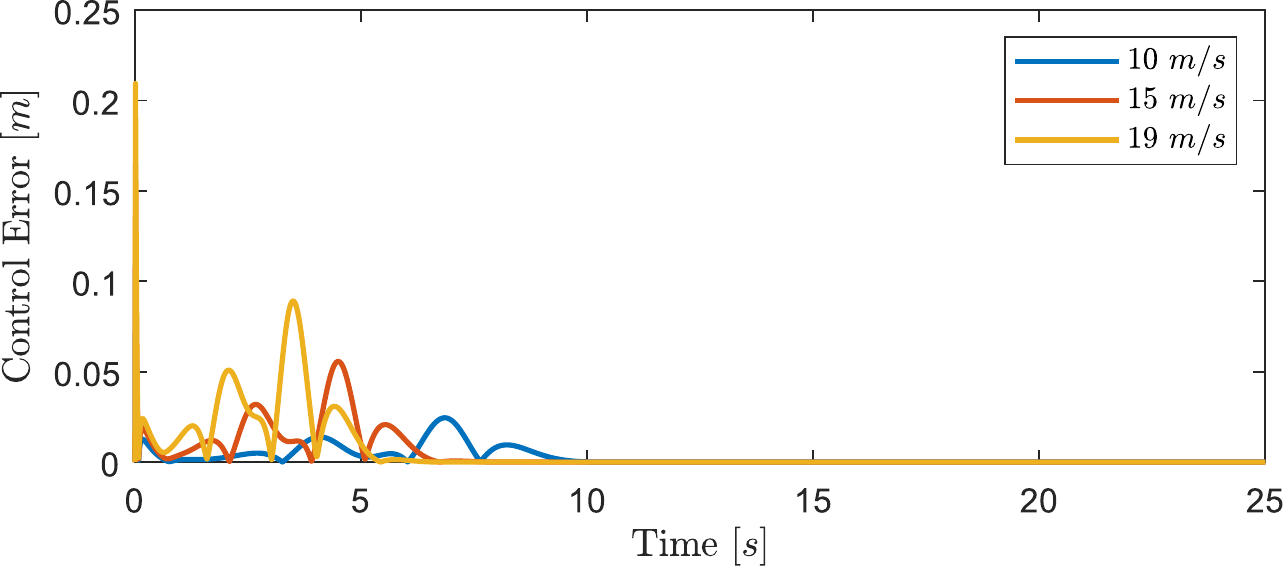}
	\vspace{-8pt}
	\captionof{figure}{{ Lane changing: control errors. Peak error (following an initial transient) under $10~cm$.}}
\end{center}
\end{figure}
% \vspace{5pt}
\begin{figure}
\begin{center} 
	\includegraphics[width=0.9\linewidth]{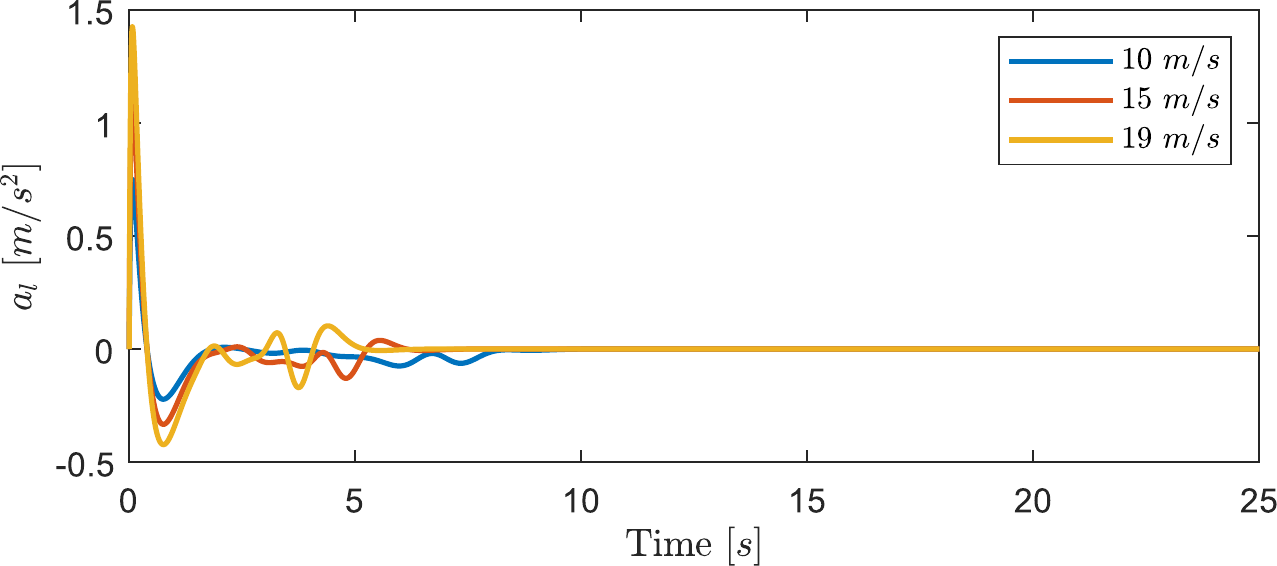}
	\vspace{-8pt}
	\captionof{figure}{{ Lane changing: longitudinal acceleration. Peak acceleration (following an initial transient) is about $0.1~m/s^2$.}}
\end{center}
\end{figure}
% \vspace{5pt}
\begin{figure}
\begin{center}
	\includegraphics[width=0.9\linewidth]{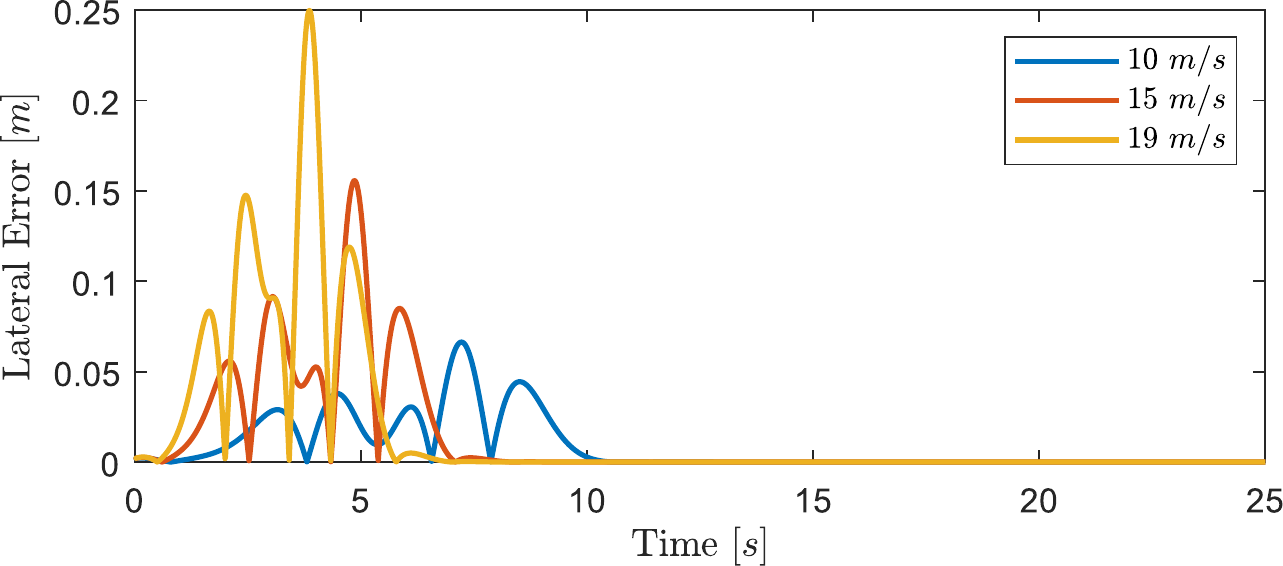}
	\vspace{-8pt}
	\captionof{figure}{{ Lane changing: lateral errors. Peak error is $25~cm$.}}
\end{center}
\end{figure}
% \vspace{5pt}
\begin{figure}
\begin{center}
	\includegraphics[width=0.9\linewidth]{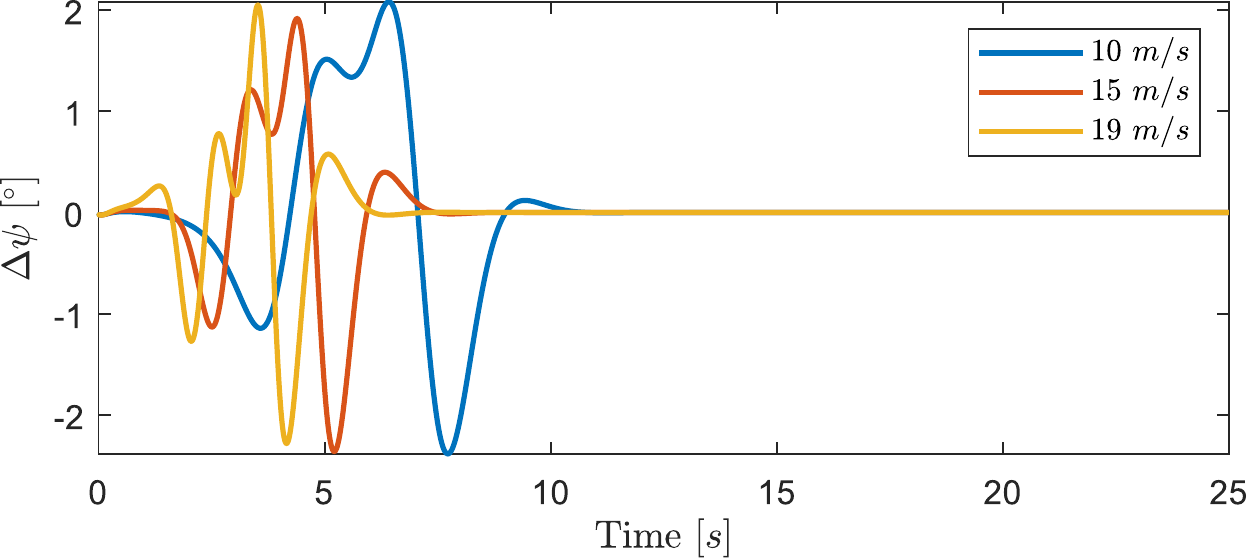}
	\vspace{-8pt}
	\captionof{figure}{{ Lane changing: heading errors. Peak error is $2.5^{\circ}$.}}
\end{center}
\end{figure}

\vspace{3pt}

\begin{center}
\begin{small}\tabcolsep=0.15cm
\begin{tabular}
{ |c| c |c | c| c|}
\hline
\multirow{2}{4em}{Speed} & \multicolumn{2}{|c|}{Peak lateral error} & \multicolumn{2}{|c|}{Peak heading error} \\
\cline{2-5}
&	Shivam 19 & Falcone 07	&	Shivam 19 & Falcone 07 \\
\hline
 $10~m/s$ &		$7~cm$ 		&	$96~cm$		&	$2.2^{\circ}$			&	$2.6^{\circ}$  \\
 $15~m/s$ & 	$16~cm$ 	&	$125~cm$	&	$2.2^{\circ}$			&	$2.67^{\circ}$ \\
 $19~m/s$ & 	$25~cm$ 	&	$158~cm$	&	$2.1^{\circ}$			&	$2.33^{\circ}$ \\
 \hline
\end{tabular}
\end{small}
\captionof{table}{\label{tab:table-two}Peak lateral and heading error for lane change}
\end{center}

Fig.~8 shows the reference and the vehicle's  trajectories for the three experiments.  The small tracking  errors at about $z_{1}=52~m$
and $z_{1}=85~m$ correspond to sharp-direction changes in the target curve and likely are due to corresponding
prediction errors.  Fig. 9 depicts the  control error, and is very small  after the lane-change maneuver is completed.
 Fig. 10 shows the longitudinal acceleration of the vehicle, which has an initial transient peak of about  $1.5~m/s^2$ for the $19~m/s$ reference  speed, and is under $0.5~m/s^2$ thereafter.
 The lateral position error, depicted in Fig. 11, has peaks of $7~cm$,
$16~cm$, $25~cm$ for the respective speeds of $10~m/s$, $15~m/s$ and      $19~m/s$. The heading errors are plotted in Fig. 12 and have a maximum peak of approximately $2.5^{\circ}$.\

Ref. \cite{Borrelli07} also provided the lateral errors and heading errors,  and the peak values (for Controller B) as well as those obtained from
our simulations are summarized in Table 2.

\section{Experimental Results}
This section presents the results of a laboratory experiment in which a platoon of four mobile robots follows a given path. The first robot tracks a point moving along the path, and  each subsequent robot tracks a point on the path relative to the position of the preceding robot. The experiment used the differential-drive robots of the Robotarium, a remotely-accessible swarm-robotics testbed at Georgia Tech \cite{Pickem17}.

\begin{figure}[b]
\centering
\begin{minipage}{0.33\linewidth}
\centering
\includegraphics[width=0.9\linewidth]{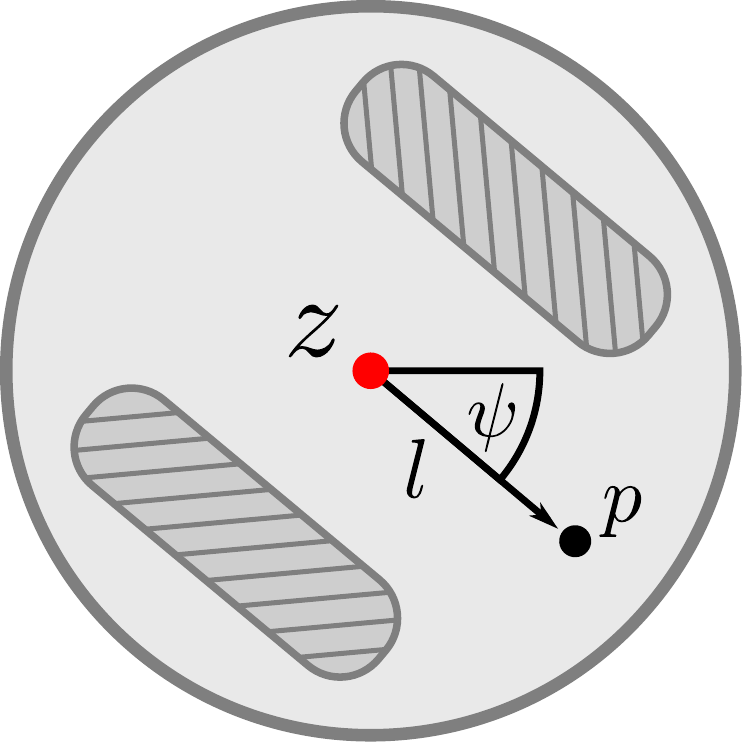}
\end{minipage}\hfill%
\begin{minipage}{0.63\linewidth}
\centering
\includegraphics[width=\linewidth]{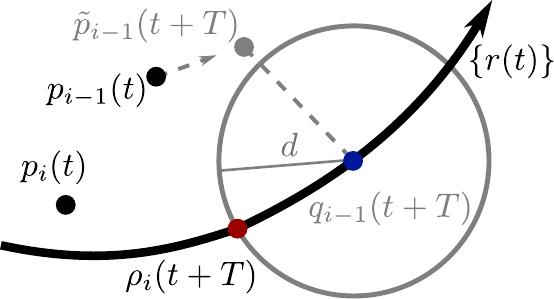}
\end{minipage}\hfill%
\caption{{Left: Unicycle and kinematic point $p$ ahead of it. Right: An illustration of Algorithm 1 for robot $i.$\label{fig:combined}}}
\end{figure}

\subsection{Robot modelling and control}
The differential-drive robots of the Robotarium are modelled by unicycle dynamics given by the following equation
\begin{equation}
\begin{pmatrix}
\dot{z}_{1}\\\dot{z}_{2}\\\dot\psi
\end{pmatrix} = \begin{pmatrix}
\cos\psi & 0\\ \sin\psi & 0 \\ 0 & 1
\end{pmatrix}\begin{pmatrix}
v_{\ell} \\ \omega
\end{pmatrix},
\label{eq:unicycle}
\end{equation}
where $z:=(z_{1},z_{2})^{\top}$ is the position of the center of gravity of the robot, $\psi$ is  its  heading,  $v_{\ell}$ is its longitudinal  velocity, and $\omega$
is its angular velocity.
In certain tracking applications one attempts to control $\{z(t)\}$ towards a curve $\{\bar{r}(t)\}$, and the control consists of the vector $(v_{\ell},\omega)^{\top}$.\footnote{The notation $\bar{p}(t)$ and not $p(t)$ will be clarified shortly, when we define the process $\{p(t)\}$.} Here, to simplify the control of the unicycle, a transformation  proposed by \cite{Olfati-Saber02} is used to map the velocity vector of a kinematic point $p:=(p_1,p_2)^\top$ at a given distance $l$ ahead of the robot    to the velocity vector $(v_{\ell},\omega)^{\top}$. This transformation
is
\begin{equation}
\left(\begin{matrix}
v_{\ell}\\ \omega
\end{matrix}\right)=\left(
\begin{matrix}
\cos\psi & \sin\psi\\
-l^{-1}\sin\psi &l^{-1}\cos\psi\\
\end{matrix}\right)\left(\begin{matrix}\dot{p}_1 \\ \dot{p}_2\end{matrix}\right);
\label{eq:nid}
\end{equation}
see Fig.~\ref{fig:combined} for a visual aid.  Under the mapping defined by Eq. \eqref{eq:nid}, it is possible to control the
robot by the state equation $\dot{p}=u$ with the input $u$, which simplifies the control law defined by Eq. (14) as compared to
what it would be in the framework of Eq. (27).

In order to comply with the notation established in Sec. 2, define $x(t):=p(t)$, then Eqs. (8) and (9) have the form
\begin{equation}
\dot{x}(t)=u(t),~~~~~~~~ y(t)=x(t).
\end{equation}
Therefore, and by Eqs. (11)-(13),
\begin{equation}
g(x(t),u(t))=x(t)+Tu(t),
\end{equation}
and hence
$\frac{\partial g}{\partial u}(x(t),u(t))=T$. Consequently, Eq. (14) defining the control law becomes
\begin{equation}
\dot{u}(t)=\frac{\alpha}{T}\big(r(t+T)-x(t)-Tu(t)\big),
\end{equation}
which does not require any numerical integration for computing $g(x(t),u(t))$ or
$\frac{\partial g}{\partial u}(x(t),u(t))$.

It should be noted that while simplifying the control law, this procedure  introduces a tracking error if the stated objective
is to have
$\{z(t)\}$, not $\{p(t)\}$  track $\{\bar{r}(t)\}$.  However, this difficulty can be rectified by defining $\{r(t)\}$ as the trajectory of $\{p(t)\}$ if $\{z(t)\}\equiv\{\bar{p}(t)\}$; $p(t)$ can be computed as the point in the plane that is $l$ meters ahead
of $\bar{p}(t)$ in the  direction $\dot{p}(t)$.
Using this transformation we consider, in the rest of this section, the control problem of having $\{p(t)\}$ track the curve $\{r(t)\}$.  

\begin{figure*}\centering
\includegraphics[width=\linewidth]{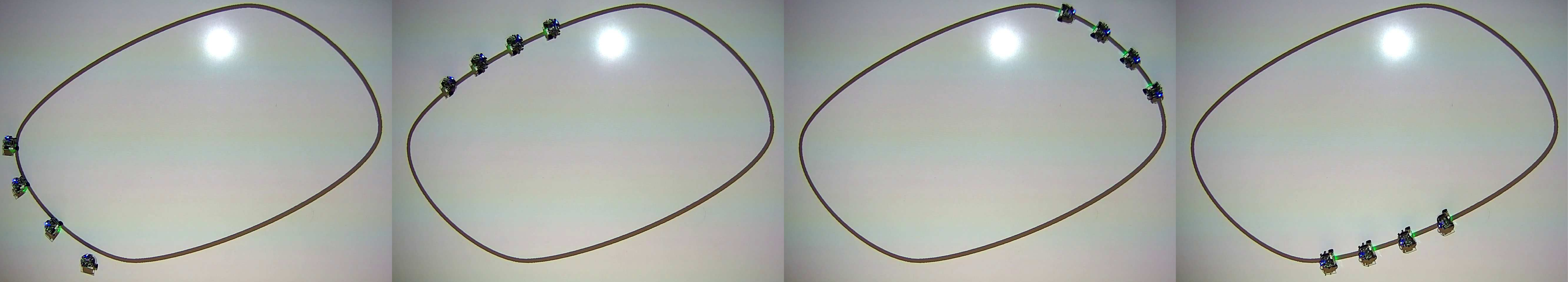}
\caption{{Stills from the Robotarium experiment. A platoon of four differential-drive robots moves along the path in a clockwise direction; the reference is displayed using an overhead projector. See \cite{video} for a video.} \label{fig:exp}}
\end{figure*}

\subsection{Tracking problem definition}
This subsection presents the tracking   algorithm used in the experiment described below. Specifically, it details the procedure for determining the reference point that 
the kinematic point $p$ of each robot must track.  The platoon is comprised of
four robots  indexed by $i=1,2,3,4$. The kinematic point associated with each robot is denoted by $p_i(t),$ and its control  is achieved as described in the previous subsection.

The kinematic point of each robot $i=1,2,3,4$ tracks a point $\rho_{i}(t+T),$ which is on the reference path $\{r(t)\}$. The kinematic point of the first robot, the platoon leader, tracks the reference point $\rho_1(t+T):=r(\gamma (t+T)),$ where $\gamma>0$ is a scaling factor chosen to limit the rate at which $\rho_1(t+T)$ moves along the path. The need for such scaling stems from the fact that, if $\gamma=1$, the robots may be  commanded to move at a higher speed than their physical limitations impose.  The reference points for each of the kinematic points of the subsequent robots are determined as follows. Let  $d>0$ be the target distance between $p_{i}$ and $p_{i-1}$, for $i=2,3,4$. At time $t$, let $\tilde{p}_{i-1}(t+T)$ be  the predicted position of the point $p_{i-1}(t)$ at time $t+T$,
and 
let $q_{i-1}(t+T)$ be the point on the curve $\{r(t)\}$ closest to  $\tilde{p}_{i-1}(t+T)$.  Let
${\tau_{i}(t+T):=\min_{\tau>t}\{||r(\tau)-q_{i}(t+T)||=d\}}$, and  set the tracking-reference point to be
${\rho_i(t+T):=r(\tau_{i-1}(t+T))}.$ This procedure is formalized below by Algorithm 1 and illustrated  in Fig.~\ref{fig:combined}, below. The expression in Eq. (31) is applied to robot $i$ by replacing $u(t)$ with $u_i(t),$ $x(t)$ with $p_i(t),$ and $r(t+T)$ with $\rho_{i}(t+T)$.

\begin{algorithm}[b]
    \caption{Reference computation for robots $i=2,3,4$.}
    \label{alg:main}
    \textbf{Inputs:} $\{r(t)\}$ and $\tilde{p}_{i-1}(t+T)$ \\ 
    \textbf{Output:} $\rho_i(t+T)$.
    \begin{algorithmic}
        \State $q_{i-1}(t+T)$ $\leftarrow$ the nearest point to  $\tilde{p}_{i-1}(t+T)$ on $\{r(\cdot)\}$
        \State $\tau_{i-1}(t+T)\leftarrow \min_{\tau>t}\{||r(\tau)-q_{i-1}(t+T)||=d\}$
        \State $\rho_{i}(t+T)\leftarrow r(\tau_{i-1}(t+T))$
    \end{algorithmic}
\end{algorithm}

A few remarks are due. First, we implicitly assume that  all of the quantities mentioned in Algorithm  1 are well defined, and expect this to be the case as long as $l\ll d,$ $d$ is  smaller than the distance travelled by the robots in $T$ seconds, and the curvature of the path and its time-derivative are not too large; these considerations guide the choice of the reference path $\{r(t)\}$ as well the parameters $\gamma,$ $l,$ $d$, and $T$ for the experiments described in the next subsection.
Second, Algorithm 1 can be performed by robots $i=2,3,4$ in a decentralized manner provided that has access to $\{r(t)\}$ and $\tilde{p}_{i-1}(t+T).$  

\subsection{Experimental Implementation}
\begin{figure}[!th]
    \centering
    \begin{minipage}{\linewidth}
    \begin{center}
         \includegraphics[width=0.9\linewidth]{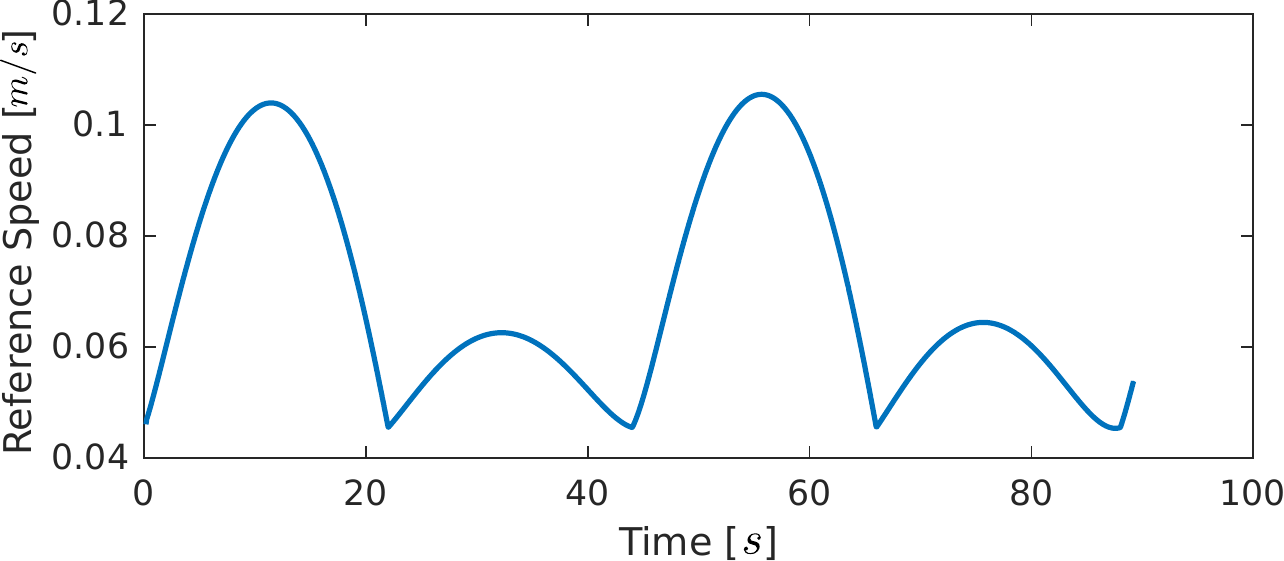}
        \vspace{-10pt}
    \captionof{figure}{Experimental setting: reference speed  \label{fig:ref-speed}}
    \end{center}    
    \end{minipage}
    \begin{minipage}{\linewidth}
    \begin{center}
    \vspace{10pt}
        \includegraphics[width=0.9\linewidth]{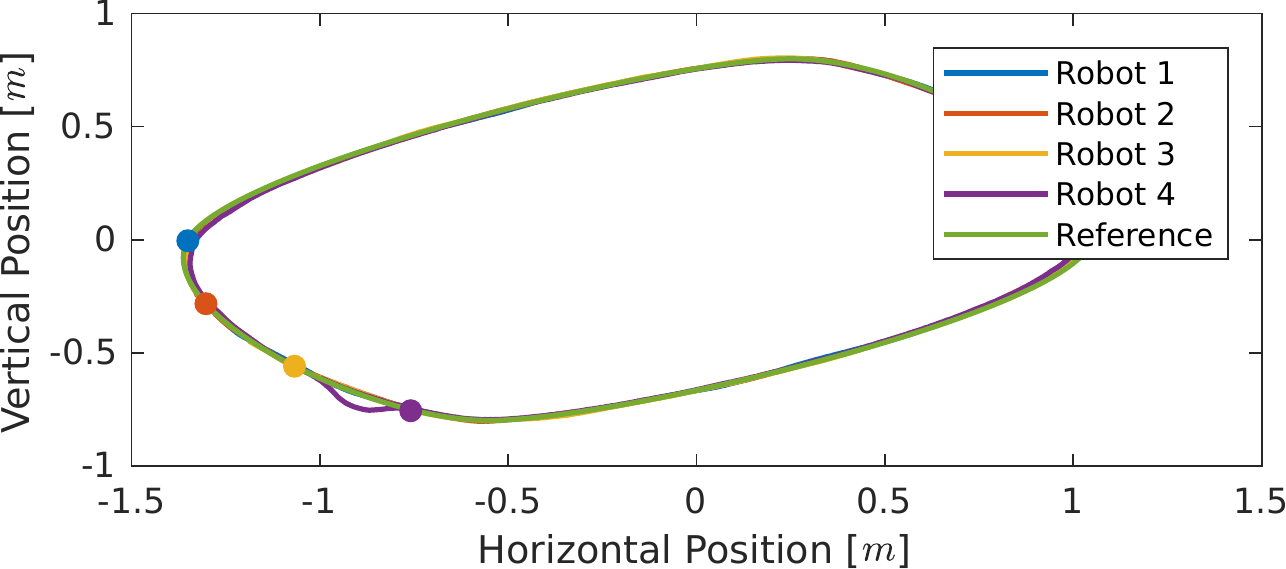}
        \vspace{-10pt}
    \captionof{figure}{ Experimental setting:   trajectories of the kinematic points \label{fig:exp-trajectories}}
    \end{center}    
    \end{minipage}
    \begin{minipage}{\linewidth}
    \begin{center}
    \vspace{10pt}
        \includegraphics[width=0.9\linewidth]{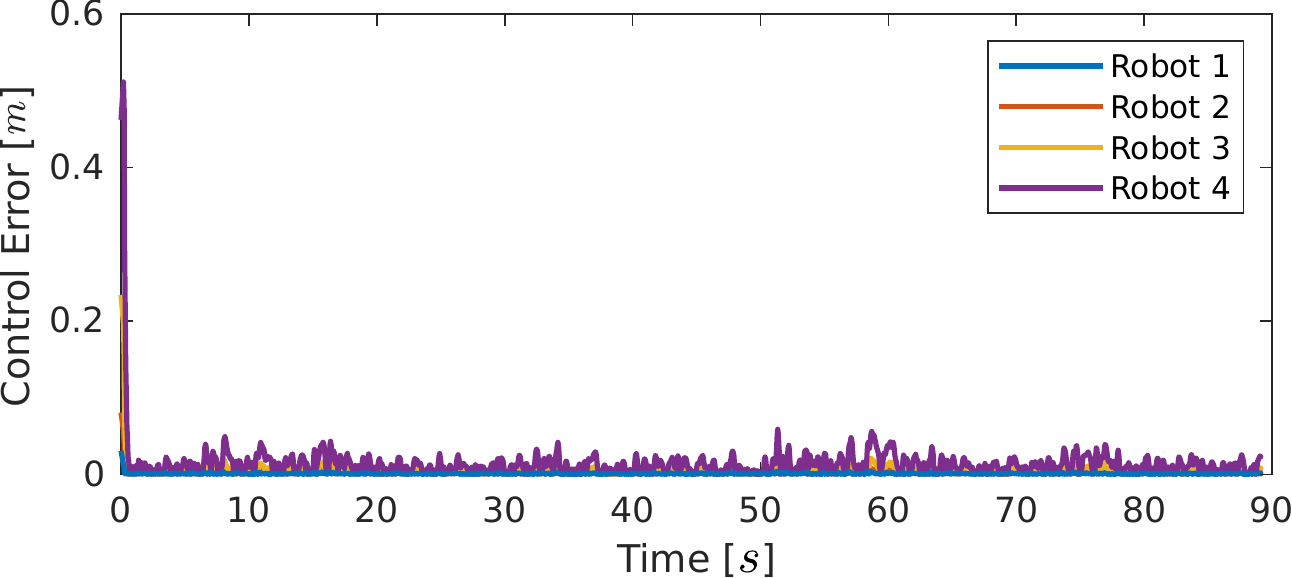}
        \vspace{-10pt}
    \captionof{figure}{ Experimental setting:  control errors  \label{fig:exp-output}}
    \end{center}    
    \end{minipage}
    \begin{minipage}{\linewidth}
    \begin{center}
    \vspace{10pt}
    %    \includegraphics[width=0.83\linewidth]{exp_tracking_error_data.pdf}
    %    \vspace{-10pt}
    %\captionof{figure}{ Experimental setting:  %tracking errors  \label{fig:exp-tracking}}
   % \end{center}    
   % \end{minipage}
   % \begin{minipage}{\linewidth}
   % \begin{center}
   % \vspace{8pt}
        \includegraphics[width=0.9\linewidth]{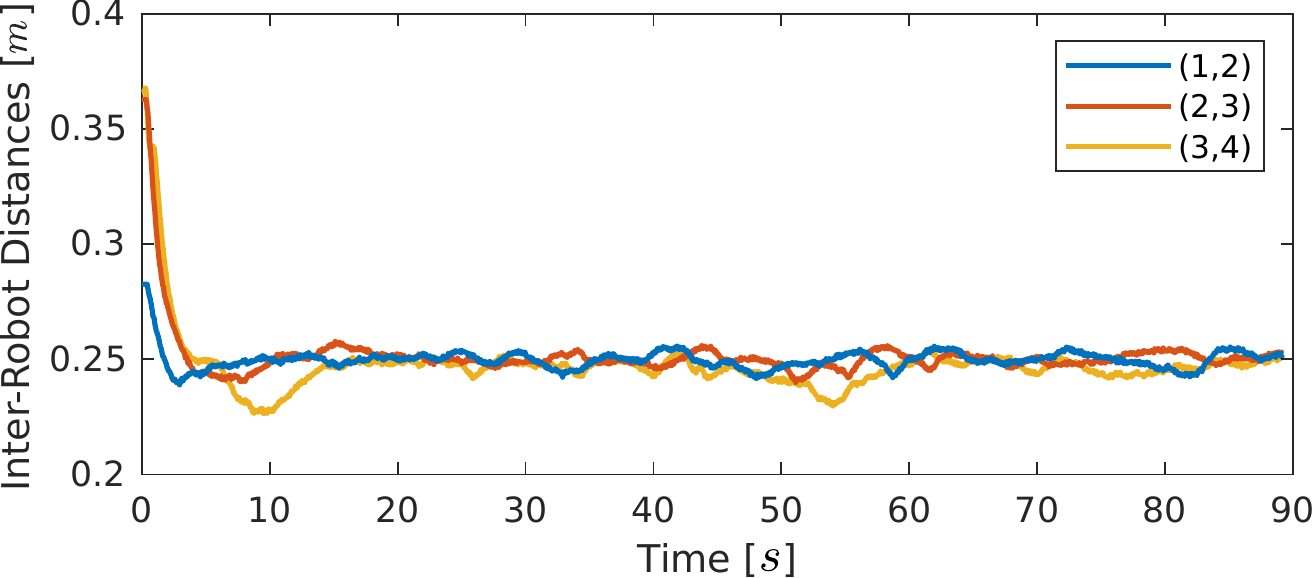}
        \vspace{-10pt}
    \captionof{figure}{ Experimental setting:  inter-spacing between kinematic points of successive robots  \label{fig:exp-distances}}
    \end{center}    
    \end{minipage}
\end{figure}

This subsection describes the laboratory experiment implemented in the Robotarium.
The reference path, displayed  on the testbed surface by an overhead projector, is shown in Fig.~\ref{fig:exp}.
The path is comprised of a  concatenation of four third-order polynomials corresponding to the four sides of the path. The speed of the target point
$r(t)$ is time-dependent, and its graph is depicted in  Fig.~\ref{fig:ref-speed}. 
The following  parameters are used:   $l=0.08~m$,   $d=0.25~m$, and   the control-algorithm's parameters are  $\alpha = 45$,  $T=0.6~s,$ and the integration step size for computing the
controls  $u_i(t)$ is $\Delta t = 0.033~s$. 
After some experimentation, we chose $\gamma = 0.0455$.

The leftmost image of Fig.~\ref{fig:exp} shows the initial positions of the robots obtained by camera, not their corresponding kinematic points.  From left to right, subsequent images show the counterclockwise progress of the robots along the path.
 Starting with the second image, near-equal distances between consecutive robots in the platoon is noted.
A more detailed view of the motion of the robots can be seen from the video clip in \cite{video}. 

Fig.~\ref{fig:exp-trajectories} shows that the trajectories of the kinematic points $p_i(t)$, whose initial positions are indicated by the dots on the path. It is evident that the points $p_{i}(t)$ closely follow the target curve after initial transients.  
The resulting control errors, i.e., 
${\|\rho_i(t+T)-\tilde{p}_i(t+T)\|}$ for  $i=1,2,3,4$,
are shown in 
Fig. \ref{fig:exp-output}.  After  ${t=1.4~s}$, the average control error for the four robots is approximately $5~mm$. 
% TODO Beef this up
% The variable speed of $\rho_1(t)$   
Finally, examining the platoon, the three distances between
successive kinematic points, $||p_{i}(t)-p_{i-1}(t)||$, $i=2,3,4$,   are shown in Fig.~\ref{fig:exp-distances}, and they seem to   approach the target distance of $25~cm$. The noted 
deviations of the last inter-spacing,
$||p_{4}(t)-p_{3}(t)||$, from the $25~cm$ target
occurs at about times $t=9.5~s$ and $t=54~s$, which
correspond to the times of highest velocities as can be seen in Fig. 15.

\section{Conclusions}
This paper presents experimental results on an output-tracking technique recently proposed by the authors. The technique is based on three principles: Newton-Raphson flow for the   solutions of time-dependent algebraic equations, output prediction, and controller speedup. Past experiments,  conducted on various systems with simple state-space models, exhibited fast tracking convergence. The main objective of this paper is to test the technique on more realistic models that are commonly used in the control of autonomous vehicles, in order to gauge its efficacy on more complicated control problems. 

The paper describes results of experiments from simulation as well as  a laboratory setting. For  the simulation studies we chose two examples of curve-tracking  problems from the literature, which  use model-predictive control
\cite{Zhou17,Borrelli07}. Comparisons of the peak lateral error and peak heading error reported in these references, versus what was reported in this paper, indicate that our results are no worse. Therefore, and since the proposed  technique does not solve differential equations and may be simpler than MPC, it may provide a framework for effective tracking control in future applications.  

The lab experiments concern a platoon of mobile robots, where the objective is to control the inter-spacing between successive vehicles (robots) to a given reference distance. The results indicate convergence.

Current research includes theoretical as well as  practical problems. The main theoretical question is how to identify conditions for stability of the closed-loop system, where a complicating factor is the fact that the controller's function  lacks a closed-form expression. A practical problem arises from the fact that currently the control technique is model-based. Current research addresses the use of  learning algorithms so as to  extend its scope  to model-free environments.
\vfill\eject

\bibliographystyle{IEEEtran}
\bibliography{Paper}

\end{document}